\documentclass[fleqn,10pt]{wlscirep}
\usepackage[utf8]{inputenc}
\usepackage[T1]{fontenc}
\usepackage{physics} 
\usepackage{xcolor}
\usepackage{graphicx}
\usepackage{multirow}
\usepackage[section]{placeins}

\title{Fabricating a dielectrophoretic microfluidic device using 3D-printed moulds and silver conductive paint}

\author[1,2]{Shayan Valijam}
\author[2]{Daniel P.G. Nilsson}
\author[2]{Dmitry Malyshev}
\author[2]{Rasmus Öberg}
\author[1]{Alireza Salehi}
\author[2,3,*]{Magnus Andersson}
\affil[1]{Faculty of Electrical Engineering, K. N. Toosi University of Technology, Tehran, 1631714191, Iran}
\affil[2]{Department of Physics, Umeå University, Umeå, 901 87, Sweden}
\affil[3]{Umeå Center for Microbial Research (UCMR), Umeå, 901 87, Sweden}

\affil[*]{magnus.andersson@umu.se}


\begin{abstract}

Dielectrophoresis is an electric field-based technique for moving neutral particles through a fluid. When used for particle separation, dielectrophoresis has many advantages compared to other methods, providing label-free operation with greater control of the separation forces. In this paper, we design, build, and test a low-voltage dielectrophoretic device using a 3D printing approach. This lab-on-a-chip device fits on a microscope glass slide and incorporates microfluidic channels for particle separation. First, we use multiphysics simulations to evaluate the separation efficiency of the prospective device and guide the design process. Second, we fabricate the device in PDMS (polydimethylsiloxane) by using 3D-printed moulds that contain patterns of the channels and electrodes. The imprint of the electrodes is then filled with silver conductive paint, making a 9 pole comb electrode. Lastly, we evaluate the separation efficiency of our device by introducing a mixture of 3 µm and 10 µm polystyrene particles and tracking their progression. Our device is able to efficiently separate these particles when the electrodes are energized with ±12 V at 75 kHz. Overall, our method allows the fabrication of cheap and effective dielectrophoretic microfluidic devices using commercial off-the-shelf equipment.

\end{abstract}
\begin{document}

\flushbottom
\maketitle

\thispagestyle{empty}

\section{Introduction}
\begin{figure}[ht]
 \centering
 \includegraphics[width=\columnwidth]{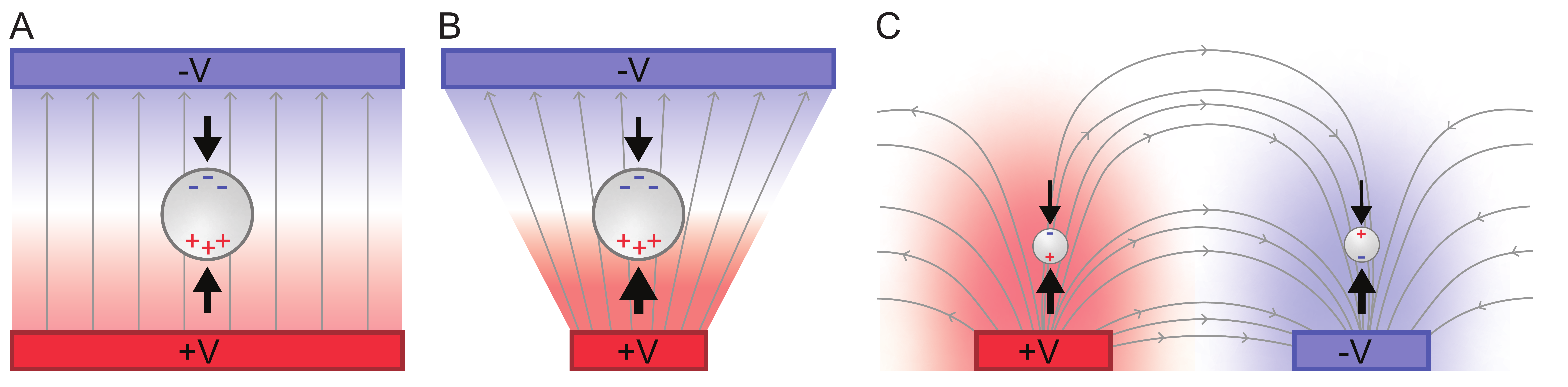} 
 \caption{The principles of dielectrophoresis on a non-conductive neutral particle. Panel A shows that a neutral particle in an alternating uniform field will have balanced forces and remain stationary. In panel B, the alternating field is instead non-uniform and the resulting net force acting on the particle will make it move. If the frequency of the field is tuned so the particle moves toward the lower field intensity it is denoted as negative DEP. Panel C shows that when the electrodes are positioned next to each other (in a comb pattern), the electric field is stronger near the electrodes resulting in a repelling force, thus pushing the particles into the center of a microfluidic channel.}
 \label{fig-1}
\end{figure}

Microfluidic devices that can separate micro-particles have important applications in a variety of fields. These include medicine, chemical engineering, wastewater treatment, environmental assessment, forensic identification, and cell separation \cite{valijam2021influence,xu2016droplet,santana2020review,verpoorte2002microfluidic}. Separation of micron-sized particles can be done using different approaches; however, for both biological and non-biological particles, dielectrophoresis (DEP) has proven to be a powerful method. The technique offers label-free, rapid, and controllable manipulation, which has been shown to perform well in both low and for high-throughput applications \cite{al2022review, ghomian2022review}. Dielectrophoresis specifically refers to the interaction between a dielectric (non-conductive) particle, and a non-uniform, alternating electric field. Consequently, the DEP force depends both on the particle characteristics, such as size, morphology, and dielectric properties \cite{velmanickam2022recent}; as well as the frequency of the electric field and how the electrodes are shaped and positioned, as shown in Fig. \ref{fig-1}. If the electrodes are positioned next to each other, the non-uniform electric field near the electrodes results in a force deflecting the particles away from their path allowing for particle separation.

Effective particle separation requires an appropriate flow rate to transport the particles. This is often achieved using microfluidic channels, which allow for controlled and laminar flow conditions. Microfluidic devices, like those used for DEP, can be fabricated using various methods such as lithography, laser photoablation, hot embossing, and direct 3D printing \cite{rogers2005recent,ferrari2022photo}. Soft lithography is of special interest, as it is relatively cheap and easy to use. This technique uses bio-compatible polymers like polydimethylsiloxane (PDMS) to fabricate microfluidic structures around premade moulds. By printing these moulds using high-precision 3D printers, such as commercial stereolithography (SLA) printers, it is possible to produce moulds with spatial resolutions down to 5 µm \cite{nilsson2022patient, subirada2022development, cheon2022fabrication, Huddy2022}. This allows us to quickly produce small channels with a high degree of precision.

One common technique to generate wide electric fields for electrokinetic particle separation is by using planar comb electrodes. However, the fabrication process of these electrodes often requires expensive techniques like physical vapour deposition (PVD), micromachining, or inkjet printing \cite{alazzam2017novel, hajari2020dielectrophoresis, li2019review, dixon2016inkjet}. Previous works have shown that it is possible to integrate the electrodes directly into the walls of a microfluidic channel using an Ag-PDMS mix; however making the nanoparticles distribute evenly throughout the PDMS has shown to be a challenge \cite{lewpiriyawong2010continuous}. A more affordable and accessible technique for creating comb electrodes is directly applying them on the microfluidic channel. By creating a 3D-printed mould for the electrodes, it is possible to simply fill the mould with conductive material, such as silver conductive paint, and join the electrodes to the microfluidic channel.

In this work, we present a method for making DEP microfluidic devices for particle separation that is quick and does not need expensive equipment or a cleanroom. We design, optimize, and evaluate our device using finite element method (FEM) simulations. We then fabricate the device by using 3D printing, PDMS moulding, and conductive silver paint to create the microfluidic channels and comb electrodes. The electrodes are energized with AC voltage. Finally, we use high-speed imaging and particle tracking to verify and evaluate the separation of 3 µm and 10 µm polystyrene beads.

\section{Results and Discussion}
\subsection{Designing the electrodes and determining their electrical parameter using simulations}
\begin{figure*}[!t] 
 \centering
 \includegraphics[width=0.9\linewidth]{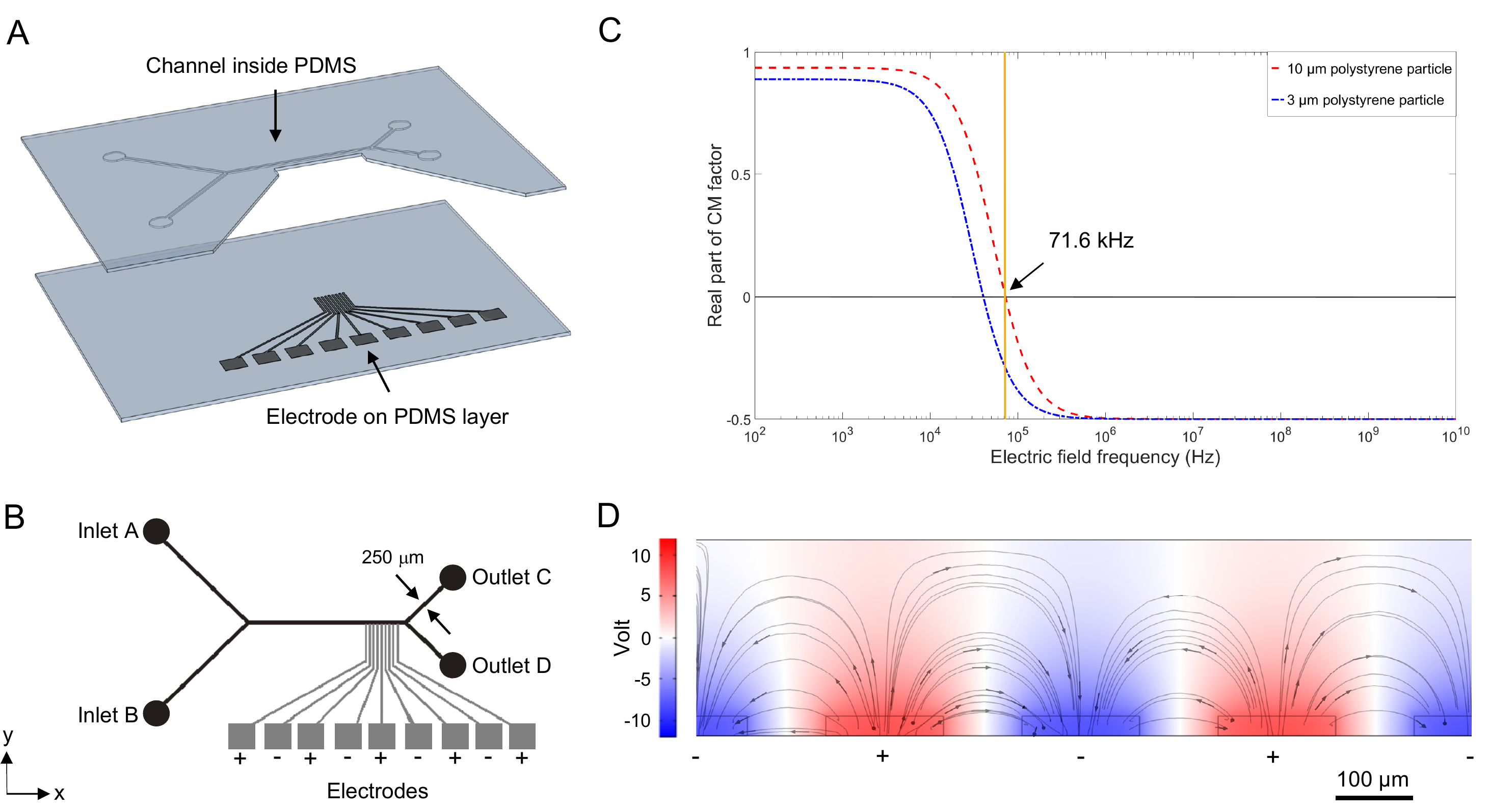}
 \caption{Panel A shows the design of the dielectrophoretic microfluidic device where the microfluidic channel (top layer) and electrodes (bottom layer) are placed on thin layers of PDMS. In panel B, we show a compounded (top) view with all the features labeled. The electrodes extend into the channel and are routed back to connection pads that are energized with AC voltage. Panel C shows the CM factor plot. The real part of the CM factor is negative for both particle sizes for frequencies above 71.6 kHz (dotted line). Panel D shows a simulation of the relative electric fields in the comb electrode, when operated at 75 kHz.}
 \label{fig-2}
\end{figure*}

To make a low-cost microfluidic device that can separate micro-sized particles using dielectrophoretic forces, we used a design-build-test approach. First, we designed the device using FEM simulations in COMSOL Multiphysics (v5.5, COMSOL AB). This allowed us to optimize the layout of the channels, electrodes, and electric parameters before constructing the physical device \cite{Valijam2023}. In our design, we took into consideration that it should be possible to produce the channels and electrodes using a 3D printer for cheap and fast fabrication. The general design of the device is shown in Fig. \ref{fig-2}A, with a detailed schematic in Fig. \ref{fig-2}B showing the layout of the inlets, electrodes, and outlets. We made the main channel 250 µm wide, 35 µm high and 10 mm long, with approach channels extending out at 45$^{\circ}$ to the in- and outlets. These allow us to introduce and extract particles from the main channel. Along the side of the main channel, nine electrodes are positioned in a comb pattern. These are 150 µm wide, placed with a center-to-center distance of 250 µm and protruding 20 µm into the flow channel. The electrodes are made flush with the channel floor and extend to a depth of 35 µm into the bottom layer. Both the top and bottom layers are made in PDMS (polydimethylsiloxane), which is a transparent, flexible and non-toxic silicone rubber. This compact design makes it possible to fit the dielectrophoretic microfluidic device on a microscope glass slide and observe the separation of biological samples in real-time.

We designed the device so that inlet A provides a controllable sheath flow through the main channel. This sheath flow focuses particles from inlet B towards the sidewall where the electrodes are positioned, enhancing the DEP-force from the electrodes felt by the particles. To generate an electric field between the electrodes, we connect them to +V and -V in an alternating pattern, see Fig. \ref{fig-2}B. Our simulations suggest that particles are efficiently focused toward the electrodes using fluid rates of 70 µL/minute and 14 µL/minute for the sheath flow and particle flow, respectively. When particles enter the separation stage (region of the nine electrodes), they will be subjected to DEP forces in the y-direction, pushing them away from the electrodes. This repulsive DEP where the particles move away from the strong electric field is commonly referred to as negative DEP (nDEP). Depending on the size and electrical properties of the particles, the nDEP force will differ in magnitude, resulting in separation between the particle species. Our aim is to separate 3 µm and 10 µm polystyrene beads into outlets C and D. These particles are approximate size analogues to platelets and red blood cells, respectively, making them relevant to biological applications. We optimize this separation by tuning the layout of the device, the electrode voltage and the driving frequency. Since the driving frequency for the electrodes determines the nDEP force, we determined the Clausius–Mossotti (CM) factor by solving Eq. \ref{eq_CM}. The CM factor is a complex number that describes the polarizability of particles and its real part for our particles is shown in Fig. \ref{fig-2}C. A negative real part of the CM factor results in a repulsive force from the electrodes on the particles, this occurs in both of our used particle sizes at frequencies above 71.6 kHz. At these frequencies, it becomes possible to separate different-sized particles using nDEP, as the nDEP force felt by the particles is proportional to the cube of the particle radius, as seen in Eq. \ref{eq_Fdep}. Increases in the driving frequency will result in a larger nDEP force; however, high frequencies are known to cause air bubble formation, and as such, we use a driving frequency of 75 kHz in our experiments \cite{lewpiriyawong2010continuous}.

\begin{figure*}[!t]
 \centering
 \includegraphics[width=1\linewidth]{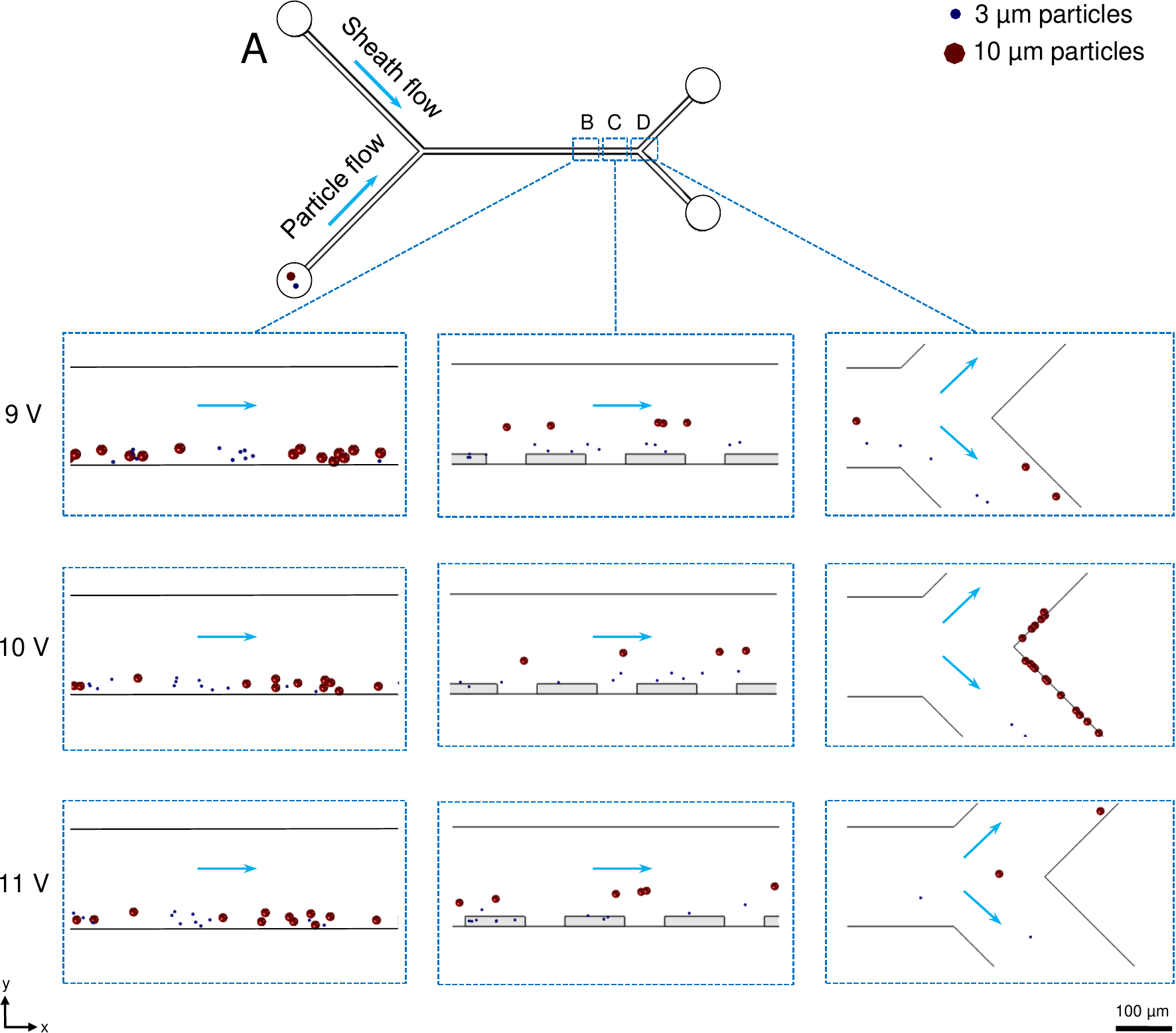}
 \caption{Numerical simulation to evaluate the voltage dependency on separation of 10 µm (red) and 3 µm (blue) particles. We test this for ±9, ±10, ±11 V and observe the particles in three key marked regions of our device: at the focusing stage (I), at the separation stage (II), at the separation junction (III). At 11 V, all particles are separated successfully according to their size.}
 \label{fig-3}
\end{figure*}

Since our aim is to make electrodes from silver conductive paint in our final device, we produced prototype electrodes to evaluate their electrical properties for use in our simulations. We found the conductivity of electrodes to be 2.5 S/m. From the simulation, we calculated the electric field strength throughout the microfluidic system to be $\sim$0.017 V/µm in the center of the channel and $\sim$0.450 V/µm close to the electrodes, with the full field distributions visualized in Fig. \ref{fig-2}D. To find the lowest voltage that yields complete separation, we simulated a particle flow through the channel while energizing the electrodes at different voltages. In this simulation, we introduced 1 ml of buffer solution containing 4 000 randomly distributed particles (2 000 of each size) at a constant flow rate. The properties of the particles and the buffer are listed in table \ref{tab:table1}. During the simulation, we monitored the particle motion in three parts of the channel as seen in Fig. \ref{fig-3}; before the separation stage (I), at the separation stage (II), and at the separation junction (III). From snapshots of these simulations for different applied voltages we found that ±9 V was not sufficient to separate the two sizes (0 \% of 10 µm particles flow through outlet C). By increasing the voltage to ±10 V, the particles are partially separated (42 \% of 10 µm particles flow through outlet C), and at ±11 V, particles are fully separated (100 \% of 10 µm particles flow through outlet C). The particle motion in these simulations can be seen in the supporting movies S1-S3, respectively. We therefore conclude that energizing the electrodes at ±11 V (75 kHz) should be sufficient for the complete separation of 3 µm and 10 µm particles.

Increases in the electric field strength result in a higher separation efficiency; however, it also causes more Joule heating. Heating the solution will change its conductivity and permittivity, and excessive heating may be harmful to biological cells \cite{matsuura2015thermodynamics}. To estimate how the temperature increases with applied voltage, we calculated the fluid temperature as
\begin{equation}
\label{eq_deltaT}
\Delta T\approx \frac{\sigma_{\text{m}}\times {V_{\text{rms}}}^2}{2K_{\text{m}}},
\end{equation}
where $V_{\text{rms}}$ is the magnitude of the applied voltage, and using $\sigma_{\text{m}}$ = 100 mS/m and $K_{\text{m}}$ = 0.60 W/m$\cdot$K for the electrical and thermal conductivity, respectively \cite{castellanos2003electrohydrodynamics}. At full separation (±11 V) we would have an increase of 10 K in the sample fluid, suggesting that voltages in this range will not significantly affect device performance or the particles.

\subsection{Building the device and testing its performance} 
\begin{figure*}[!htb]
 \centering
 \includegraphics[width=0.85\linewidth]{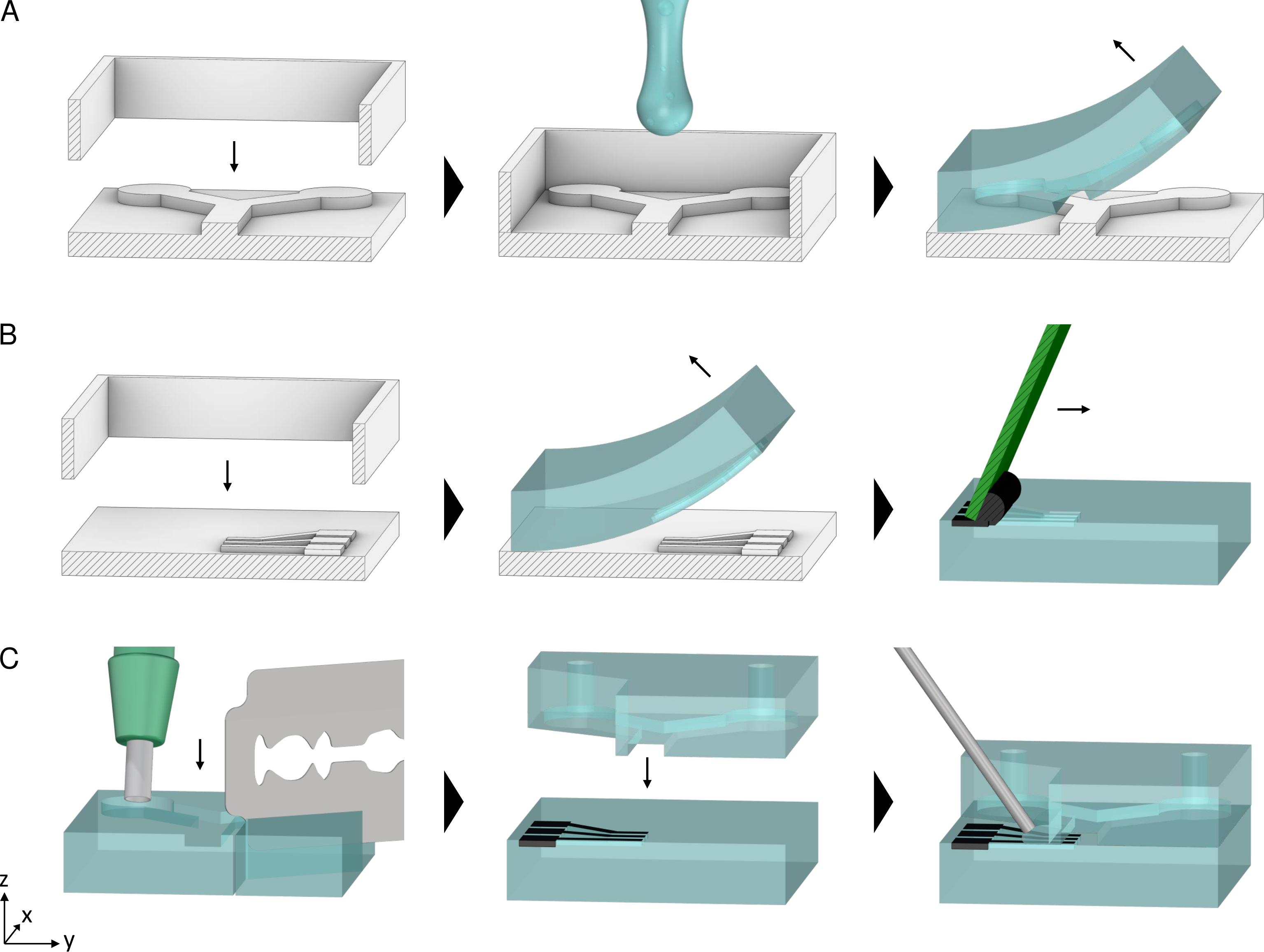} 
 \caption{The main steps in the procedure of making a nDEP microfluidic device. For visualization purposes, the illustration is not to scale and cut in half along x-axis. In panel A, we show how the top layer of the device is made. PDMS is poured into a 3D-printed mould (and its frame). Once cured, the PDMS is peeled off with an imprint of the flow channel. In panel B, a similar procedure for the bottom layer is shown. The imprint for the electrodes is filled with silver conductive paint after the PDMS is cured. In panel C, a biopsy punch is used to make openings in the channel before both halves are bonded together. To improve the bonding along the electrodes, we cut the corresponding part of the top layer away and added uncured PDMS at the seam using a syringe.}
 \label{fig-4}
\end{figure*}

\begin{figure*}[!htb] 
 \centering
 \includegraphics[width=1\linewidth]{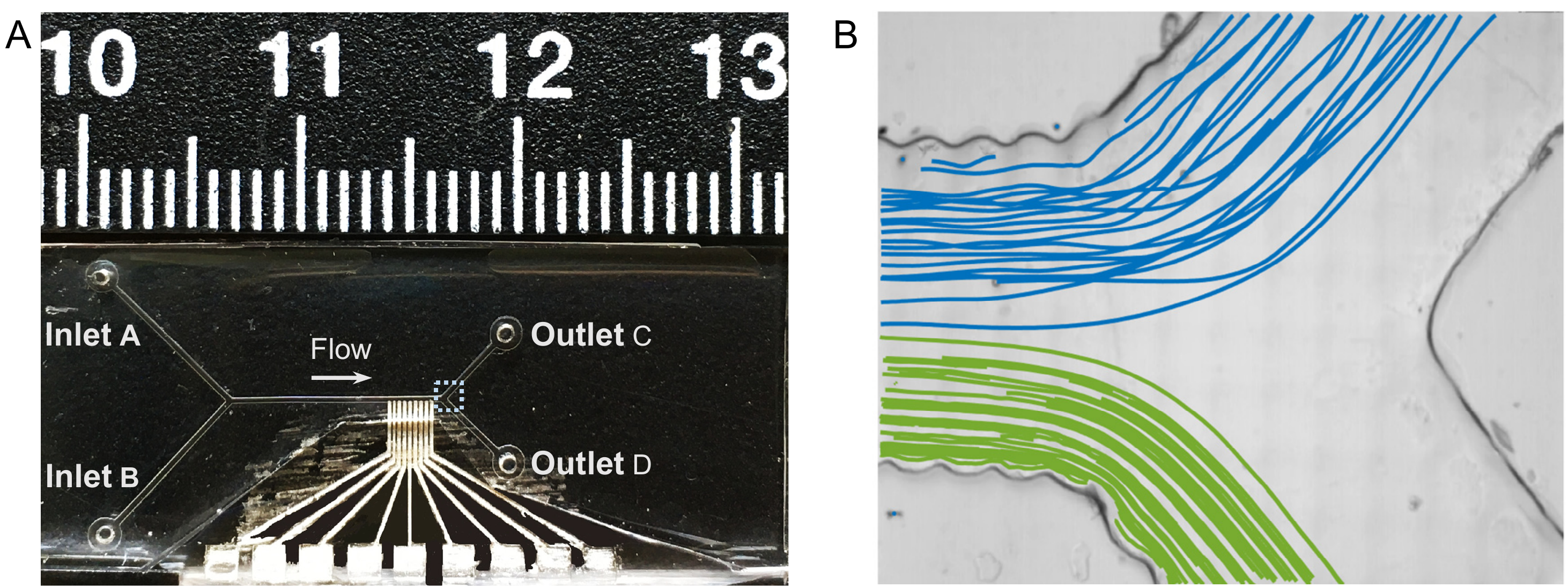} 
 \caption{Panel A shows our microfluidic chip with the electrodes visible and a cm scale bar above. The separation junction is marked with a dashed box in Panel A and is shown in high resolution in Panel B. Particle traces are shown for 3 µm (green) and 10 µm (blue) particles when energizing the electrodes at ±12 V. Particle traces before the separation stage and at the separation stage are shown in Figure S1 Panel A and B, respectively.}
 \label{fig-5}
\end{figure*}

Using the results from our simulation, we built a lab-on-a-chip device using a 3D printing and PDMS moulding approach, similar to Ho \textit{et al.,} 2015 \cite{ho20153d}. First, the design of the simulated device was exported to our SLA resin printer. Using this printer, we produced a high-resolution (50 µm) printout of the channel, onto which 1 mm of PDMS was moulded to make the microfluidic layer of the device, as seen in Fig. \ref{fig-4}A. Second, a similar PDMS layer was made with imprints of the electrodes and these were filled with silver conductive paint, as seen in Fig. \ref{fig-4}B. Third, holes were made in the microfluidic layer for the in/outlets and then the two layers were bonded together, as seen in Fig. \ref{fig-4}C. Finally, nine copper wires were connected between the electrode pads and a signal generator (Model 8102, Topward), set to operate at ±11 V and 75 kHz. The manufacturing process is described in greater detail in Sec. \ref{subsec:DeviceFabrication}, and a photo of the finished device is shown in Fig. \ref{fig-5}A. To allow fluid flow through the device, we added tubing to the inlets and outlets and connected the device to a syringe pump (VIT-FIT, LAMBDA Laboratory Instruments). The injection rates for inlets A and B were set to a continuous 70 µL/minute and 14 µL/minute as suggested by simulations. 

To inlet B, we connected a syringe containing a phosphate-buffered saline (PBS) solution (0.1 S/m) and a mixture of 3 and 10 µm beads, as defined in Sec. \ref{subsec:SamplePreparation}. The device was then placed in an optical microscope (Microphot-FX, Nikon Instruments) with a 20X objective (427958, Nikon Instruments) for tracking particles through the channel. Due to the high particle velocities, we mounted a high-speed CCD camera (MotionBLITZ EoSens Cube7, Mikrotron) and set it to record videos at 1696x1240 pixel resolution and 735 frames per second. Particle trajectories were analyzed from the video using the free tracking software, ToxTrac (v2.98) \cite{rodriguez2018toxtrac,rodriguez2017toxid}. We first analysed region I (before the separation stage) to see if the flow ratio from the simulations would focus the particles to the side where the electrodes were positioned. Indeed, we observed that particles were focused closer to this side and that particles moved with the flow and were not exposed to external forces, see Movie S4 and Figure S1A. Next, we analysed region II (separation stage) and region III (junction). Initially, when using the same voltage as in the simulation (±11 V), we observed that particle separation was not satisfactory. We therefore increased the voltage to ±12 V, at which full separation was achieved. Since the electric field is the cause of Joule heating, we recalculated the temperature increase using Eq. \ref{eq_deltaT}. At a separation voltage of ±12 V the total temperature increase is 12 K, which is only 2 K higher than at ±11 V, and still low enough to not affect the device or the sample significantly.

Running the device at ±12 V, we analysed the trajectories of the larger (10 µm) particles at the separation stage (electrodes 4 to 7) and found that these particles had a noticeable change in their y-directions, see Movie S5 and Figure S1B. The gradient of the 10 µm particles was assessed to 0.042±0.0013, by fitting a linear regression model to the trajectories. This implies that the 10 µm particles moved about 6.3 µm in the y-direction over a 150 µm distance (equal to the width of an electrode).

After this, we monitored the separation junction. At the separation junction, we observed that particles were clearly separated by size, see Movie S6. To quantify this, we used Toxtrac and set the software to track particles according to size. Trajectories of about 30 particles from each group are shown in Fig. \ref{fig-5}B, in which we have plotted 3 µm and 10 µm particles in green and blue, respectively.

When imaging the microfluidic channels, we noted a wave-like pattern along the channel walls. These waves are caused by exposure bleeding (light leakage from LCD pixels) during 3D printing of the moulds and can be mitigated by tuning the printer and using resin additives. To ensure this pattern does not affect particle separation efficiency, we simulated the impact of these artifacts, as seen in Figure S2. The results show that the wave-like pattern does not affect the particle trajectory as compared to channels with straight walls. Rather, only subtle changes to the flow profile close to the channel walls were observed. This is similar to what we observed in experiments, where we noted that particles that move close to the surface follow the streamlines of the wall. However, these streamlines do not significantly impact the particle separation efficiency.

\FloatBarrier

\subsection{Evaluating the microfluidic channels and electrodes}
\begin{figure*}[!htb] 
 \centering
 \includegraphics[width=1\linewidth]{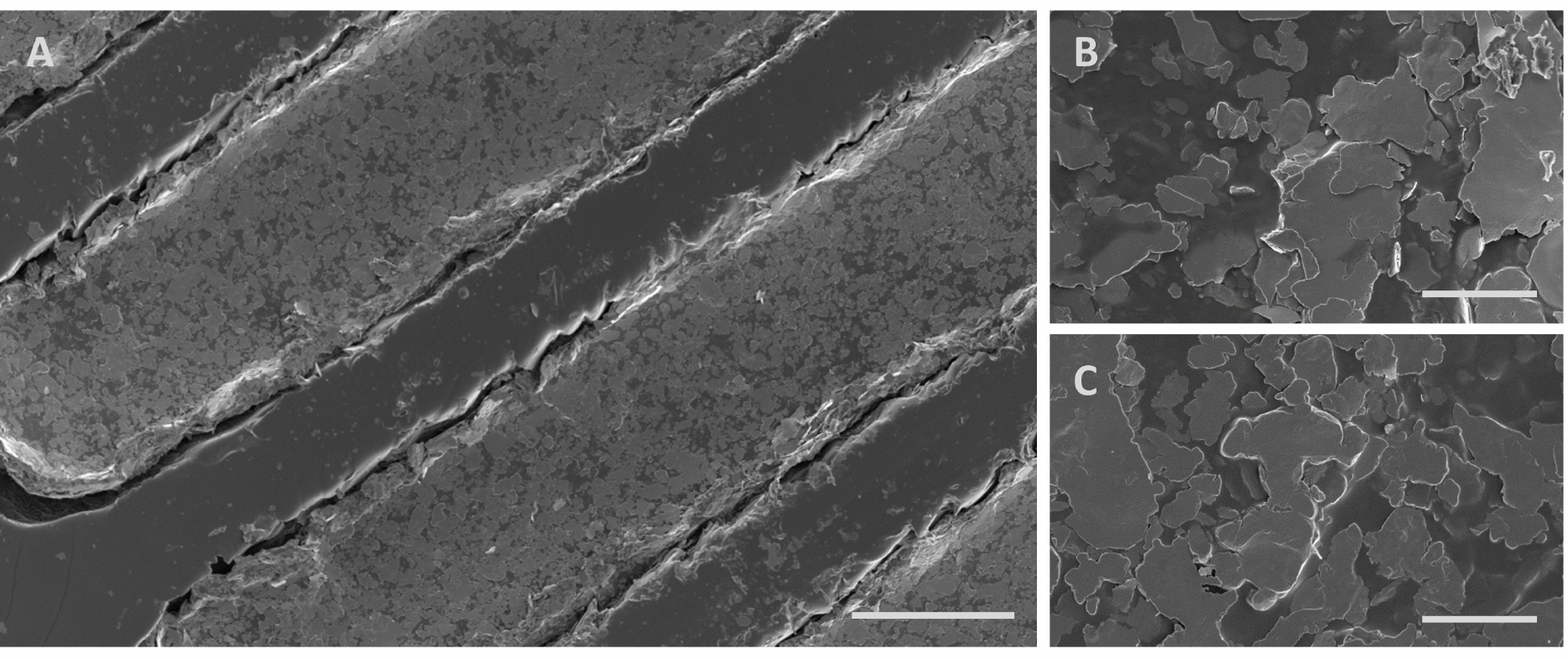} 
 \caption{EM micrographs of the comb electrode after stress-testing with 15 V for 1 hour, followed by immersion in a buffer solution and heating to 100 °C. In panel A, the electrodes are still showing an intact structure (scale bar is 100 µm). The microstructure of the electrodes at 5 000$\times$ magnification is shown, before (B) and after (C) exposure, with no visual differences (scale bars are 10 µm).}
 \label{fig-6}
\end{figure*}

To evaluate the robustness of the silver conductive paint electrodes, we tested their durability and observed their surface morphology. To test their durability, we energized the electrodes at a higher voltage than what was used during the experiments, ±15 V compared to ±12 V, at the same frequency of 75 kHz. The electrodes were also immersed in a buffer solution for 1 h and heated to 100 $^\circ$C. After these tests, we imaged the electrodes using SEM to evaluate their surface integrity, as shown in Fig. \ref{fig-6}. We found that none of the three stress tests affected the surface morphology of the painted electrodes compared to controls, either on the macro-structure as seen in Fig. \ref{fig-6}A, or on the microstructure as seen in Fig. \ref{fig-6}B-C. Thus, we concluded that silver conductive paint electrodes are sufficiently durable for the purpose of our dielectrophoretic device.

\FloatBarrier

\section{Conclusion}
In this study we developed a lab-on-a-chip device that utilizes dielectrophoresis to sort uncharged microparticles based on their size using 3D-printed microfluidic channels equipped with silver conductive paint electrodes. We designed our device with FEM simulations to optimise the electrodes' frequency and voltage for sorting 3 µm and 10 µm plastic particles and we built the device in PDMS using 3D-printed moulds that contain imprints of the channels and electrodes. We evaluated the performance of the chip using particle tracking, showing our device could successfully separate 10 µm from 3 µm particles. Thus, our proposed method allows for the fabrication of cheap dielectrophoretic microfluidic devices using off-the-shelf equipment, that can be used in various sorting applications, for example, sorting cells, wastewater treatment, and separating bacterial spores from environmental samples for detection.

\section{Materials and Methods}

\subsection{Theoretical background}
Dielectric (non-conductive) particles become polarized when placed in a non-uniform electric field. When polarised, these particles experience a force, the magnitude of which depends on the polarisability of the particle \cite{Grimnes2015}. This is the dielectrophoresis (DEP) force, as shown in Fig. \ref{fig-1}, and differences in this force can be used to redirect and sort particles. For homogenous spherical particles, the DEP force is given by
\begin{equation}
\label{eq_Fdep}
F_{\text{DEP}}=2\pi r^3\varepsilon_{\text{0}}\varepsilon_{\text{m}}\Re[CM(f)]\div\abs{\Vec{E}}^2,
\end{equation}

\noindent
where \(r\) is the radius of the particle, while \(\varepsilon_{\text{m}}\) and \(\varepsilon_{\text{0}}\) are the permittivity of the suspension medium and vacuum, respectively. The gradient of the electric field is given by \(\div\abs{\Vec{E}}^2\). Finally, Re\([CM(f)]\) represents is the real part of the Clausius-Mossotti \((CM)\) factor. The \(CM\) factor can vary between -0.5 and 1.0 and is dependent on the particle's complex permittivity and thus the operating frequency $f$ of the applied electric field. This factor also determines what direction the particle will move \cite{AbdRahman2017}. If \(CM\) is positive, the DEP force is towards the stronger electric field, referred to as pDEP. If the \(CM\) factor is negative, the particle moves away from the stronger electric field, and is called nDEP \cite{Salimi2019}. The device we use in this study works with nDEP, as shown in Fig. \ref{fig-2}C. The \(CM\) factor is quantified as

\begin{equation}
\label{eq_CM}
CM=\frac{\varepsilon_{\text{p}}^*-\varepsilon_{\text{m}}^*}{\varepsilon_{\text{p}}^*+2\varepsilon_{\text{m}}^*}.
\end{equation}

\noindent where the complex permittivity of the particle and medium are given by \(\varepsilon_{\text{p}}^*\) and \(\varepsilon_{\text{m}}^*\) respectively. Thus the \(CM\) factor value can be between -0.5 (\(\varepsilon_{\text{p}}^*\) $<<$ \(\varepsilon_{\text{m}}^*\)) and 1 (\(\varepsilon_{\text{p}}^*\) $>>$ \(\varepsilon_{\text{m}}^*\)). These two parameters are defined as 

\begin{equation}
	\label{eq_epsilon}
	\left\{
    \begin{array}{lll}
      \varepsilon_{\text{p}}^*= \varepsilon_{\text{p}}-j\frac{\sigma_{\text{p}}}{w}, \\
      \varepsilon_{\text{m}}^*= \varepsilon_{\text{m}}-j\frac{\sigma_{\text{m}}}{w}, \\
    \end{array}
    \right.
\end{equation}

\noindent respectively, where the conductivity of the particle and medium are \(\sigma_{\text{p}}\) and \(\sigma_{\text{m}}\), respectively. Further, \(w=2\pi f\) is the angular frequency and \(j = \sqrt{-1}\). 

The other major force affecting the movement of the suspended particles in microfluidic devices, is the hydrodynamic drag force. Hydrodynamic drag force for spherical particles in laminar flow is calculated from Stokes' law

\begin{equation}
\label{eq_Fdrag}
F_{\text{Drag}}=6\pi r_{\text{ext}}\eta(\Vec{u}-\Vec{u}_\text{p}),
\end{equation}

\noindent where \(r_{\text{ext}}\) is the external radius of the particle, \(\eta\) is the viscosity of the medium, \(\Vec{u}\) and \(\Vec{u}_{\text{p}}\) are the flow velocity and the particle velocity respectively. We defined the physical and electrical properties of the particles and the medium as described in the literature \cite{khoshmanesh2010size, zhu2009dielectrophoretic}, and the parameters used in our simulations are listed in Table \ref{tab:table1}.

\subsection{Numerical Simulation}
We simulated the device using a three dimensional model in COMSOL Multiphysics (v5.5, COMSOL AB). We used the laminar flow module to solve the Navier-Stokes equation and obtain the flow field. The AC/DC module was used to simulate non-uniform electric field by solving Laplace’s equation. The fluid (water) was assumed incompressible, and the flow was assumed to be laminar and steady. Both these are reasonable assumptions for flows at these low Reynolds numbers. A non-slip boundary condition was applied to the channels, and the pressure at all inlets and outlets was kept zero. A constant flow was assumed for the inlets, with the injection flow rate for inlet A and B set to 70 µL/min and 14 µL/min respectively. Newton's second law was employed to solve the particle trajectory, given by

\begin{equation}
	\label{eq_NII}
	\left\{
    \begin{array}{lll}
      \nabla{p}+\eta\laplacian{\Vec{u}}+\Vec{F}_{\text{e}}=0, \\
      \laplacian{u}=0, \\
      \div\Vec{u}=0, \\
    \end{array}
    \right.
\end{equation}

\noindent where \(P\) and \(\Vec{F}_{\text{e}}\) are the pressure and applied forces, respectively. The properties of the suspension medium were set to those of water, with density 1000 kg/m$^3$, dynamic viscosity 1.0$\times10^{-3}$ Pa$\cdot$s, and relative permittivity 80 \cite{jia2015continuous}. The Laplace equation was solved to investigate applied electrical potential

\begin{equation}
\label{eq_pot}
\laplacian{\varphi}=0,
\end{equation}

\noindent where \(\varphi\) is the electric potential applied to the electrodes, while all other boundaries were considered insulating.
By solving Eq. \ref{eq_Fdep} and Eq. \ref{eq_epsilon}, the position of the particles were obtained and calculated using

\begin{equation}
\label{eq_pos}
{\vec{F}_{\text{DEP}}}+{\vec{F}_{\text{drag}}}=m\dfrac{d\vec{u}_{\text{p}}}{dx}
\end{equation}

\noindent where \(m\) and \(t\) are  the mass of the particle and time, respectively. Our simulation model counted 38 886 mesh elements in total.

\begin{table}[h]
\centering
\caption{Physical and dielectric properties of the particles and suspension medium.}
\label{tab:table1}
\begin{tabular}{llll}
\hline
\multicolumn{1}{c}{\multirow{2}{*}{Property}} & \multicolumn{2}{c}{Particle} & \multicolumn{1}{c}{\multirow{2}{*}{Medium}} \\ \cline{2-3}
\multicolumn{1}{c}{} & 10 µm & 3 µm & \multicolumn{1}{c}{} \\ \hline
Conductivity (S/m) & 2.5e-2 & 4.5e-2 & 0.1 \\
Relative permittivity & 5.0$\times\varepsilon_{\text{0}}$ & 2.56$\times\varepsilon_{\text{0}}$ & 80$\times\varepsilon_{\text{0}}$ \\
Dynamic viscosity (Pa$\cdot$s) & - & - & 1.0e-3 \\
Density (kg/m$^3$) & - & - & 1.0e+3 \\ \hline
\end{tabular}
\end{table}

\subsection{Device fabrication}
\label{subsec:DeviceFabrication}
We manufactured our device using 3D printing and PDMS moulding. Our design consists of two halves, a microfluidic half, and a half containing the electrodes, as seen in Fig. \ref{fig-4}. 3D moulds for both these parts were made, and then used to create the device geometry in PDMS (Sylgard 184, Dow Corning). To make the moulds quick and cheap to manufacture, we used a desktop SLA 3D printer (Photon Mono SE, Anycubic). The geometry of the device was designed in COMSOL Multiphysics and prepared for the printer using a slicer software (Photon Workshop V2.1.29, Anycubic). For printing, the channels and electrodes were inverted (protruding) and printed with a 1 mm thick backing that will form the bottom of the mould. To ensure accurate dimensions of the features, print settings had to be adjusted in the slicer software. Settings of layer thickness and exposure times are dependent on both the printer and resin. This calibration was done first using the coverslip method, as described in \cite{Nilsson2022}, and then by printing open-source calibration models.

Before we could start moulding, the printed parts had to be treated to make them inert, as failing to do so would inhibit the curing of the PDMS. This procedure consisted of; first, cleaning the part in isopropanol for 3 min using a sonicator (Super RK 100, Bandelin Sonorex). Second, curing the part in a UV box (15 W at 400 ± 10 nm FWHM) for 2 h. And lastly, degassing the part in an oven at 70 $^\circ$C for 6 h. When the part was ready, it was clamped down (facing up) with a frame to complete the mould. A mixture of 10:1 PDMS base and curing agent was then poured into the mould to a thickness of 1 mm and degassed with a vacuum desiccator. It was then placed in an oven at 70 $^\circ$C for 1 h to cure. The PDMS layer containing the imprint was finished and could be peeled off. This process was repeated for both parts of the device. In the top part, through-holes were made for in and outlets to the flow channel using a 1 mm biopsy punch (Integra Miltex).

In the bottom part; to make the electrodes flush with the channel floor, we filled the imprints in the PDMS layer with 3 µl silver conductive paint (SCP03B, Electrolube), scraping off the excess. The paint consists of 45$\%$ silver particles, has a viscosity of 70 mPa$\cdot$s (at 25$^\circ$C) and dries in $\sim$10 minutes. Both of the PDMS layers were then treated in \(O_{\text{2}}\) plasma for 1 min (0.5 mbar chamber pressure) using a plasma cleaner (Atto, Deiner Electronic), before being aligned and bonded together under an optical microscope. Finally, the device was placed on a hotplate for 5 min (at 90$^\circ$C) to improve the adhesion. However, it was found that some fluid could leak out where the electrodes bonded to the top layer. We therefore cut away the part of the top layer in contact with the electrodes and rebonded the two halves. Uncured PDMS was then poured on the exposed electrodes and cured, sealing the joint. To apply a voltage to the electrodes, we connected 9 copper wires to the extended silver pads. The channels were then washed with deionized water (for 2 min) ahead of use.

\subsection{Sample Preparation}
\label{subsec:SamplePreparation}
To test the performance of the device, we prepared a mixture of 3 ± 0.09 µm (C37484, Invitrogen CML Latex, Thermo Fisher Scientific) and 10 ± 1.0 µm (C37259, Invitrogen CML Latex, Thermo Fisher Scientific) latex polystyrene beads. The stock volume for 10 and 3 µm particles are 4.1 and 4.3 g/mL, respectively, and were diluted 3000 times with deionized water to achieve appropriate concentrations. To achieve a medium better suited for nDEP, we modified the electrical conductivity of the particle and sheath flow medium to 0.1 S/m, by adding 1X phosphate-buffered saline (PBS) solution at a volume ratio of 1:100.

\subsection{Particle tracking}
To track the particles in our microchannel, we used the free image tracking software called ToxTrac (v2.98) \cite{rodriguez2018toxtrac}. Videos were first preprocessed in UMUTracker \cite{zhang2017umutracker} to remove the background and binarize the videos. The videos were loaded in Toxtrac and we defined the region that we wanted to track as an arena. In Toxtrac, we could set the tracking algorithm to track particles depending on size. Since some particles had low contrast and were hard to track between frames, we used the multitracking ID-algorithm to have higher stability in the trajectories \cite{rodriguez2017toxid}. Despite this, some trajectories were broken. However, this did not affect the final results since we could merge broken tracks and set the tracking algorithm to identify particles based on their sizes when they exit the device. Using this approach and monitoring the whole separation junction, we could identify the size of particles when they exit either outlet A or B.

To calculate the rate of change of particles in the y-direction in the separation region, we fitted a linear regression model to individual trajectories. We multiplied this number with the pixel to µm conversion factor (2.5 pixel/µm), defined by the sensor's pixel size (8 µm) and microscope magnification (x20).

\subsection{SEM Imaging}
\noindent
We used SEM to image the silver conductive paint electrodes after different treatments. To prevent charge buildup on the PDMS surface, we coated the sample with a 5 nm thick layer of platinum using a sputter coater (Q150T-ES, Quorum Technologies Ltd). We then imaged the electrodes using a scanning electron microscope (Merlin FESEM, Carl Zeiss), with its InLens imaging mode at a magnifications of 100-50000$\times$.

\section*{Acknowledgements}
This work was supported by the Swedish Research Council (2019-04016); the Swedish Foundation for Strategic Research; the Umeå University Industrial Doctoral School (IDS); Kempestiftelserna (JCK-1916.2); Swedish Department of Defence, Project no. 470-A400821.

The authors acknowledge the facilities and technical assistance of the Umeå Core Facility for Electron Microscopy (UCEM) at the Chemical Biological Centre (KBC), Umeå University, a part of the National Microscopy Infrastructure NMI (VR-RFI 2016-00968).

\section*{Author contributions statement}

M.A. and S.V. conceived the experiments. S.V. and D.N conducted the dielectrophoresis experiments. D.M. did SEM imaging. D.N., S.V. and R.Ö. manufactured the electrodes. S.V. and D.N created the models used in the manuscript. D.N, D.M.and M.A. analysed the results. All authors reviewed and edited the manuscript.

\section*{Competing interests}
The authors declare that they have no known competing financial interests or personal relationships that could have appeared to influence the work reported in this paper.

\section*{Data availability statement}
The datasets generated during the current study are available from the corresponding author on reasonable request.

\bibliography{reference}

\begin{thebibliography}{10}
\urlstyle{rm}
\expandafter\ifx\csname url\endcsname\relax
  \def\url#1{\texttt{#1}}\fi
\expandafter\ifx\csname urlprefix\endcsname\relax\def\urlprefix{URL }\fi
\expandafter\ifx\csname doiprefix\endcsname\relax\def\doiprefix{DOI: }\fi
\providecommand{\bibinfo}[2]{#2}
\providecommand{\eprint}[2][]{\url{#2}}

\bibitem{valijam2021influence}
\bibinfo{author}{Valijam, S.} \& \bibinfo{author}{Salehi, A.}
\newblock \bibinfo{journal}{\bibinfo{title}{Influence of the obstacles on
  dielectrophoresis-assisted separation in microfluidic devices for cancerous
  cells}}.
\newblock {\emph{\JournalTitle{Journal of the Brazilian Society of Mechanical
  Sciences and Engineering}}} \textbf{\bibinfo{volume}{43}},
  \bibinfo{pages}{1--14} (\bibinfo{year}{2021}).

\bibitem{xu2016droplet}
\bibinfo{author}{Xu, L.}, \bibinfo{author}{Peng, J.}, \bibinfo{author}{Yan,
  M.}, \bibinfo{author}{Zhang, D.} \& \bibinfo{author}{Shen, A.~Q.}
\newblock \bibinfo{journal}{\bibinfo{title}{Droplet synthesis of silver
  nanoparticles by a microfluidic device}}.
\newblock {\emph{\JournalTitle{Chemical Engineering and Processing: Process
  Intensification}}} \textbf{\bibinfo{volume}{102}}, \bibinfo{pages}{186--193}
  (\bibinfo{year}{2016}).

\bibitem{santana2020review}
\bibinfo{author}{Santana, H.}, \bibinfo{author}{Silva, J.},
  \bibinfo{author}{Aghel, B.} \& \bibinfo{author}{Ortega-Casanova, J.}
\newblock \bibinfo{journal}{\bibinfo{title}{Review on microfluidic device
  applications for fluids separation and water treatment processes}}.
\newblock {\emph{\JournalTitle{SN Applied Sciences}}}
  \textbf{\bibinfo{volume}{2}}, \bibinfo{pages}{1--19} (\bibinfo{year}{2020}).

\bibitem{verpoorte2002microfluidic}
\bibinfo{author}{Verpoorte, E.}
\newblock \bibinfo{journal}{\bibinfo{title}{Microfluidic chips for clinical and
  forensic analysis}}.
\newblock {\emph{\JournalTitle{Electrophoresis}}}
  \textbf{\bibinfo{volume}{23}}, \bibinfo{pages}{677--712}
  (\bibinfo{year}{2002}).

\bibitem{al2022review}
\bibinfo{author}{Al-Ali, A.}, \bibinfo{author}{Waheed, W.},
  \bibinfo{author}{Abu-Nada, E.} \& \bibinfo{author}{Alazzam, A.}
\newblock \bibinfo{journal}{\bibinfo{title}{A review of active and passive
  hybrid systems based on dielectrophoresis for the manipulation of
  microparticles}}.
\newblock {\emph{\JournalTitle{Journal of Chromatography A}}}
  \bibinfo{pages}{463268} (\bibinfo{year}{2022}).

\bibitem{ghomian2022review}
\bibinfo{author}{Ghomian, T.} \& \bibinfo{author}{Hihath, J.}
\newblock \bibinfo{journal}{\bibinfo{title}{Review of dielectrophoretic
  manipulation of micro and nanomaterials: Fundamentals, recent developments,
  and challenges}}.
\newblock {\emph{\JournalTitle{IEEE Transactions on Biomedical Engineering}}}
  (\bibinfo{year}{2022}).

\bibitem{velmanickam2022recent}
\bibinfo{author}{Velmanickam, L.}, \bibinfo{author}{Jayasooriya, V.},
  \bibinfo{author}{Vemuri, M.~S.}, \bibinfo{author}{Tida, U.~R.} \&
  \bibinfo{author}{Nawarathna, D.}
\newblock \bibinfo{journal}{\bibinfo{title}{Recent advances in
  dielectrophoresis toward biomarker detection: A summary of studies published
  between 2014 and 2021}}.
\newblock {\emph{\JournalTitle{Electrophoresis}}}
  \textbf{\bibinfo{volume}{43}}, \bibinfo{pages}{212--231}
  (\bibinfo{year}{2022}).

\bibitem{rogers2005recent}
\bibinfo{author}{Rogers, J.~A.} \& \bibinfo{author}{Nuzzo, R.~G.}
\newblock \bibinfo{journal}{\bibinfo{title}{Recent progress in soft
  lithography}}.
\newblock {\emph{\JournalTitle{Materials today}}} \textbf{\bibinfo{volume}{8}},
  \bibinfo{pages}{50--56} (\bibinfo{year}{2005}).

\bibitem{ferrari2022photo}
\bibinfo{author}{Ferrari, E.}, \bibinfo{author}{Nebuloni, F.},
  \bibinfo{author}{Rasponi, M.} \& \bibinfo{author}{Occhetta, P.}
\newblock \bibinfo{title}{Photo and soft lithography for organ-on-chip
  applications}.
\newblock In \emph{\bibinfo{booktitle}{Organ-on-a-Chip}},
  \bibinfo{pages}{1--19} (\bibinfo{publisher}{Springer}, \bibinfo{year}{2022}).

\bibitem{nilsson2022patient}
\bibinfo{author}{Nilsson, D.~P.} \emph{et~al.}
\newblock \bibinfo{journal}{\bibinfo{title}{Patient-specific brain arteries
  molded as a flexible phantom model using 3d printed water-soluble resin}}.
\newblock {\emph{\JournalTitle{Scientific reports}}}
  \textbf{\bibinfo{volume}{12}}, \bibinfo{pages}{1--9} (\bibinfo{year}{2022}).

\bibitem{subirada2022development}
\bibinfo{author}{Subirada, F.} \emph{et~al.}
\newblock \bibinfo{journal}{\bibinfo{title}{Development of a custom-made 3d
  printing protocol with commercial resins for manufacturing microfluidic
  devices}}.
\newblock {\emph{\JournalTitle{Polymers}}} \textbf{\bibinfo{volume}{14}},
  \bibinfo{pages}{2955} (\bibinfo{year}{2022}).

\bibitem{cheon2022fabrication}
\bibinfo{author}{Cheon, J.} \& \bibinfo{author}{Kim, S.}
\newblock \bibinfo{journal}{\bibinfo{title}{Fabrication and demonstration of a
  3d-printing/pdms integrated microfluidic device}}.
\newblock {\emph{\JournalTitle{Recent Progress in Materials}}}
  \textbf{\bibinfo{volume}{4}}, \bibinfo{pages}{1--1} (\bibinfo{year}{2022}).

\bibitem{Huddy2022}
\bibinfo{author}{Huddy, J.~E.} \& \bibinfo{author}{Scheideler, W.~J.}
\newblock \bibinfo{journal}{\bibinfo{title}{{Protocol for deposition of
  conductive oxides onto 3D-printed materials for electronic device
  applications}}}.
\newblock {\emph{\JournalTitle{STAR Protocols}}} \textbf{\bibinfo{volume}{3}},
  \bibinfo{pages}{101523}, \doiprefix\url{10.1016/j.xpro.2022.101523}
  (\bibinfo{year}{2022}).

\bibitem{alazzam2017novel}
\bibinfo{author}{Alazzam, A.}, \bibinfo{author}{Mathew, B.} \&
  \bibinfo{author}{Alhammadi, F.}
\newblock \bibinfo{journal}{\bibinfo{title}{Novel microfluidic device for the
  continuous separation of cancer cells using dielectrophoresis}}.
\newblock {\emph{\JournalTitle{Journal of separation science}}}
  \textbf{\bibinfo{volume}{40}}, \bibinfo{pages}{1193--1200}
  (\bibinfo{year}{2017}).

\bibitem{hajari2020dielectrophoresis}
\bibinfo{author}{Hajari, M.}, \bibinfo{author}{Ebadi, A.},
  \bibinfo{author}{Farshchi~Heydari, M.~J.}, \bibinfo{author}{Fathipour, M.} \&
  \bibinfo{author}{Soltani, M.}
\newblock \bibinfo{journal}{\bibinfo{title}{Dielectrophoresis-based
  microfluidic platform to sort micro-particles in continuous flow}}.
\newblock {\emph{\JournalTitle{Microsystem Technologies}}}
  \textbf{\bibinfo{volume}{26}}, \bibinfo{pages}{751--763}
  (\bibinfo{year}{2020}).

\bibitem{li2019review}
\bibinfo{author}{Li, Q.} \emph{et~al.}
\newblock \bibinfo{journal}{\bibinfo{title}{Review of printed electrodes for
  flexible devices}}.
\newblock {\emph{\JournalTitle{Frontiers in Materials}}}
  \textbf{\bibinfo{volume}{5}}, \bibinfo{pages}{77} (\bibinfo{year}{2019}).

\bibitem{dixon2016inkjet}
\bibinfo{author}{Dixon, C.}, \bibinfo{author}{Ng, A.~H.},
  \bibinfo{author}{Fobel, R.}, \bibinfo{author}{Miltenburg, M.~B.} \&
  \bibinfo{author}{Wheeler, A.~R.}
\newblock \bibinfo{journal}{\bibinfo{title}{An inkjet printed, roll-coated
  digital microfluidic device for inexpensive, miniaturized diagnostic
  assays}}.
\newblock {\emph{\JournalTitle{Lab on a Chip}}} \textbf{\bibinfo{volume}{16}},
  \bibinfo{pages}{4560--4568} (\bibinfo{year}{2016}).

\bibitem{lewpiriyawong2010continuous}
\bibinfo{author}{Lewpiriyawong, N.}, \bibinfo{author}{Yang, C.} \&
  \bibinfo{author}{Lam, Y.~C.}
\newblock \bibinfo{journal}{\bibinfo{title}{Continuous sorting and separation
  of microparticles by size using ac dielectrophoresis in a pdms microfluidic
  device with 3-d conducting pdms composite electrodes}}.
\newblock {\emph{\JournalTitle{Electrophoresis}}}
  \textbf{\bibinfo{volume}{31}}, \bibinfo{pages}{2622--2631}
  (\bibinfo{year}{2010}).

\bibitem{Valijam2023}
\bibinfo{author}{Valijam, S.}, \bibinfo{author}{Salehi, A.} \&
  \bibinfo{author}{Andersson, M.}
\newblock \bibinfo{journal}{\bibinfo{title}{{Design of a low-voltage
  dielectrophoresis lab-on-the chip to separate tumor and blood cells}}}.
\newblock {\emph{\JournalTitle{Microfluidics and Nanofluidics}}}
  \textbf{\bibinfo{volume}{27}} (\bibinfo{year}{2023}).

\bibitem{matsuura2015thermodynamics}
\bibinfo{author}{Matsuura, Y.} \emph{et~al.}
\newblock \bibinfo{journal}{\bibinfo{title}{Thermodynamics of protein
  denaturation at temperatures over 100 c: Cuta1 mutant proteins substituted
  with hydrophobic and charged residues}}.
\newblock {\emph{\JournalTitle{Scientific reports}}}
  \textbf{\bibinfo{volume}{5}}, \bibinfo{pages}{1--9} (\bibinfo{year}{2015}).

\bibitem{castellanos2003electrohydrodynamics}
\bibinfo{author}{Castellanos, A.}, \bibinfo{author}{Ramos, A.},
  \bibinfo{author}{Gonzalez, A.}, \bibinfo{author}{Green, N.~G.} \&
  \bibinfo{author}{Morgan, H.}
\newblock \bibinfo{journal}{\bibinfo{title}{Electrohydrodynamics and
  dielectrophoresis in microsystems: scaling laws}}.
\newblock {\emph{\JournalTitle{Journal of Physics D: Applied Physics}}}
  \textbf{\bibinfo{volume}{36}}, \bibinfo{pages}{2584} (\bibinfo{year}{2003}).

\bibitem{ho20153d}
\bibinfo{author}{Ho, C. M.~B.}, \bibinfo{author}{Ng, S.~H.},
  \bibinfo{author}{Li, K. H.~H.} \& \bibinfo{author}{Yoon, Y.-J.}
\newblock \bibinfo{journal}{\bibinfo{title}{3d printed microfluidics for
  biological applications}}.
\newblock {\emph{\JournalTitle{Lab on a Chip}}} \textbf{\bibinfo{volume}{15}},
  \bibinfo{pages}{3627--3637} (\bibinfo{year}{2015}).

\bibitem{rodriguez2018toxtrac}
\bibinfo{author}{Rodriguez, A.} \emph{et~al.}
\newblock \bibinfo{journal}{\bibinfo{title}{Toxtrac: a fast and robust software
  for tracking organisms}}.
\newblock {\emph{\JournalTitle{Methods in Ecology and Evolution}}}
  \textbf{\bibinfo{volume}{9}}, \bibinfo{pages}{460--464}
  (\bibinfo{year}{2018}).

\bibitem{rodriguez2017toxid}
\bibinfo{author}{Rodriguez, A.}, \bibinfo{author}{Zhang, H.},
  \bibinfo{author}{Klaminder, J.}, \bibinfo{author}{Brodin, T.} \&
  \bibinfo{author}{Andersson, M.}
\newblock \bibinfo{journal}{\bibinfo{title}{Toxid: an efficient algorithm to
  solve occlusions when tracking multiple animals}}.
\newblock {\emph{\JournalTitle{Scientific reports}}}
  \textbf{\bibinfo{volume}{7}}, \bibinfo{pages}{1--8} (\bibinfo{year}{2017}).

\bibitem{Grimnes2015}
\bibinfo{author}{Grimnes, S.} \& \bibinfo{author}{Martinsen, {\O}.~G.}
\newblock \bibinfo{title}{{Electrodes}}.
\newblock In \emph{\bibinfo{booktitle}{Bioimpedance and Bioelectricity
  Basics}}, \bibinfo{pages}{179--254},
  \doiprefix\url{10.1016/B978-0-12-411470-8.00007-6}
  (\bibinfo{publisher}{Elsevier}, \bibinfo{year}{2015}).

\bibitem{AbdRahman2017}
\bibinfo{author}{{Abd Rahman}, N.}, \bibinfo{author}{Ibrahim, F.} \&
  \bibinfo{author}{Yafouz, B.}
\newblock \bibinfo{journal}{\bibinfo{title}{{Dielectrophoresis for Biomedical
  Sciences Applications: A Review}}}.
\newblock {\emph{\JournalTitle{Sensors}}} \textbf{\bibinfo{volume}{17}},
  \bibinfo{pages}{449}, \doiprefix\url{10.3390/s17030449}
  (\bibinfo{year}{2017}).

\bibitem{Salimi2019}
\bibinfo{author}{Salimi, E.} \& \bibinfo{author}{Bridges, G.}
\newblock \bibinfo{title}{{Dielectric Properties of Cells}}.
\newblock In \emph{\bibinfo{booktitle}{Comprehensive Biotechnology}},
  \bibinfo{pages}{585--598}, \doiprefix\url{10.1016/B978-0-444-64046-8.00061-6}
  (\bibinfo{publisher}{Elsevier}, \bibinfo{year}{2019}).

\bibitem{khoshmanesh2010size}
\bibinfo{author}{Khoshmanesh, K.} \emph{et~al.}
\newblock \bibinfo{journal}{\bibinfo{title}{Size based separation of
  microparticles using a dielectrophoretic activated system}}.
\newblock {\emph{\JournalTitle{Journal of applied physics}}}
  \textbf{\bibinfo{volume}{108}}, \bibinfo{pages}{034904}
  (\bibinfo{year}{2010}).

\bibitem{zhu2009dielectrophoretic}
\bibinfo{author}{Zhu, J.} \& \bibinfo{author}{Xuan, X.}
\newblock \bibinfo{journal}{\bibinfo{title}{Dielectrophoretic focusing of
  particles in a microchannel constriction using dc-biased ac flectric
  fields}}.
\newblock {\emph{\JournalTitle{Electrophoresis}}}
  \textbf{\bibinfo{volume}{30}}, \bibinfo{pages}{2668--2675}
  (\bibinfo{year}{2009}).

\bibitem{jia2015continuous}
\bibinfo{author}{Jia, Y.}, \bibinfo{author}{Ren, Y.} \& \bibinfo{author}{Jiang,
  H.}
\newblock \bibinfo{journal}{\bibinfo{title}{Continuous dielectrophoretic
  particle separation using a microfluidic device with 3d electrodes and
  vaulted obstacles}}.
\newblock {\emph{\JournalTitle{Electrophoresis}}}
  \textbf{\bibinfo{volume}{36}}, \bibinfo{pages}{1744--1753}
  (\bibinfo{year}{2015}).

\bibitem{Nilsson2022}
\bibinfo{author}{Nilsson, D. P.~G.} \emph{et~al.}
\newblock \bibinfo{journal}{\bibinfo{title}{{Patient-specific brain arteries
  molded as a flexible phantom model using 3D printed water-soluble resin}}}.
\newblock {\emph{\JournalTitle{Scientific Reports}}}
  \textbf{\bibinfo{volume}{12}}, \bibinfo{pages}{10172},
  \doiprefix\url{10.1038/s41598-022-14279-7} (\bibinfo{year}{2022}).

\bibitem{zhang2017umutracker}
\bibinfo{author}{Zhang, H.}, \bibinfo{author}{Stangner, T.},
  \bibinfo{author}{Wiklund, K.}, \bibinfo{author}{Rodriguez, A.} \&
  \bibinfo{author}{Andersson, M.}
\newblock \bibinfo{journal}{\bibinfo{title}{Umutracker: A versatile matlab
  program for automated particle tracking of 2d light microscopy or 3d digital
  holography data}}.
\newblock {\emph{\JournalTitle{Computer Physics Communications}}}
  \textbf{\bibinfo{volume}{219}}, \bibinfo{pages}{390--399}
  (\bibinfo{year}{2017}).

\end{thebibliography}

\end{document}